\documentclass[iop]{emulateapj}
\usepackage{color}
\usepackage{graphicx}
\usepackage{amsmath}
\usepackage{color}
\usepackage[section]{placeins}

\shortauthors{Geng et al.}

\begin{document}

\title{Gravitational-wave constraints on the cosmic opacity at $z\sim 5$: forecast from space
gravitational-wave antenna DECIGO }

\author{Shuaibo Geng\altaffilmark{1}, Shuo Cao\altaffilmark{1$\ast$}, Tonghua Liu\altaffilmark{1$\dag$}, Marek Biesiada\altaffilmark{2}, Jingzhao Qi\altaffilmark{3}, Yuting Liu\altaffilmark{1}, and Zong-Hong Zhu\altaffilmark{1}}

\altaffiltext{1}{Department of Astronomy, Beijing Normal University,
Beijing 100875, China; \emph{caoshuo@bnu.edu.cn};
\emph{liutongh@mail.bnu.edu.cn}; } \altaffiltext{2}{National Centre
for Nuclear Research, Pasteura 7, 02-093 Warsaw, Poland}
\altaffiltext{3}{Department of Physics, College of Sciences,
Northeastern University, 110819 Shenyang, China}

\begin{abstract}

Since gravitational waves (GWs) propagate freely through a
perfect fluid, coalescing compact binary systems as standard sirens
allow to measure the luminosity distance directly and provide
distance measurements unaffected by the cosmic opacity. DECi-hertz
Interferometer Gravitational-wave Observatory (DECIGO) is a future
Japanese space gravitational-wave antenna sensitive to frequency
range between target frequencies of LISA and ground-based detectors.
Combining the predicted future GW observations from DECIGO and three
current popular astrophysical probes (HII regions, SNe Ia Pantheon
sample, quasar sample) in electromagnetic (EM) domains, one would be
able to probe the opacity of the Universe at different redshifts. In
this paper, we show that the cosmic opacity parameter can be
constrained to a high precision ($\Delta \epsilon\sim 10^{-2}$) out
to high redshifts ($z\sim$5). In order to reconstruct the evolution
of cosmic opacity without assuming any particular functional form of
it, the cosmic opacity tests should be applied to individual
redshift bins independently. Therefore, we also calculate the
optical depth at individual redshifts and averaged $\tau(z)$ within
redshift bins. Our findings indicate that, compared with the results
obtained from the HII galaxies and Pantheon SNe Ia, there is an
improvement in precision when the quasar sample is considered. While
non-zero optical depth is statistically significant only for
redshift ranges $0<z<0.5$, $1<z<2$, and $2.5<z<3.5$,  such tendency
is different from that obtained in the framework of its parametrized
form. Therefore the importance of cosmic-opacity test without a
prescribed phenomenological function should be emphasized.

\end{abstract}


\maketitle


\section{Introduction}

One of the directly observed evidences for Universe undergoing an
accelerated expansion at present stage is dimming of type Ia
supernovae (SNe Ia) \citep{Riess98,Perlmutter99}. The new component
with a negative pressure called dark energy has been proposed to
explain the dimming of SNe Ia
\citep{Ratra88,Caldwell98,Cao11a,Cao13a,Cao14,Ma17,Qi18} and its
nature still remains the biggest problem in cosmology and
fundamental physics. This is not the only mechanism which
contributes to SNe Ia dimming. Standard astrophysical effects
comprise photon absorption or scattering by dust particles in the
Milky Way, intervening galaxies or the host galaxy
\citep{Tolman30,Menard10a,Menard10b,Xie15,Vavrycuk19}. All of them
have been taken into account according to the best of our current
knowledge in the preparation of the data sets we used.

In general, any effect that causes the loss or
non-conservation of photon number in the beam can contribute to the
dimming of distant objects like SNe Ia. Common name for such
non-standard or yet unknown mechanisms is cosmic opacity. Some
exotic ideas related to cosmic opacity are, for example:  conversion
of photons into light axions \citep{Csaki02,Avgoustidis10,Jaeckel10}
or gravitons \citep{Chen95} in the presence of extragalactic
magnetic fields, Kaluza-Klein modes associated with extra-dimensions
\citep{Deffayet00}. Up to now, more than 1000 SNe Ia have been
detected \citep{Scolnic18} and there are the systematic errors
rather than statistical ones, which dominate when uses SNe Ia to
constrain cosmological parameters. Cosmic opacity might be one
source of such systematic errors. Therefore, in the era of precision
cosmology it is necessary to accurately quantify any relevant
dimming effects.

In the past, the opacity of the universe has been widely
investigated by using various astronomical observations
\citep{More09,Nair12,Chen12,Li13,Holanda13,Liao13,Liao15,Jesus17,Wang17}.
These works generally fall into two categories. The first was to
combine the opacity-free angular diameter distances (ADDs) inferred
from baryon acoustic oscillations (BAO) or galaxy clusters
\citep{More09,Nair12,Chen12,Li13} with the luminosity distances
(LDs) derived from SNe Ia observations (opacity-dependent). It
should be stressed that this approach relies on the so-called
``distance duality relation'' (DDR)
\citep{Etherington1,Etherington2,Cao11}. The DDR holds in all
cosmological models described by Riemannian geometry and states that
LD and ADD should satisfy the relation $D_L=D_A(1+z)^2$, where
$D_\mathrm{L}$ and $D_\mathrm{A}$ are respectively the LD and ADD at
the same redshift $z$. Any deviation from the DDR can contribute to
the non-conservation of the photon number \citep{Ellis}. Hence,
exploring the DDR is equivalent to testing the opacity of the
universe, and most of the previous analyses based on currently
available $D_A$ data \citep{Cao17a,Cao17b} still indicated
negligible opacity of the Universe. It is obvious that cosmic
opacity assessed from the DDR is frequency independent, in contrast
to most physically viable mechanisms of opacity affecting $D_L$
measurements. We recommend a detailed discussion of
\citet{Vavrycuk20} concerning cosmic opacity tests based on the DDR.
We will come back to this issue in the concluding section.

In the second approach, a cosmological model independent method was
proposed by confronting the opacity-independent luminosity distances
inferred from the Hubble parameter $H(z)$ measurements of
differential ages of passively evolving  galaxies (which represent
cosmological standard clocks)
\citep{Holanda13,Liao13,Liao15,Jesus17,Wang17,Liu20a} with the
(opacity-dependent) luminosity distances of SNe Ia. If universe is
opaque, the flux from a standard candle received by the observer
will be reduced by a factor $e^{-\tau(z)}$, where $\tau > 0$ is the
opacity parameter which represents the optical depth associated to
the cosmic absorption.

As an alternative to the above mentioned methods, we will use
gravitational waves (GWs) as standard sirens \citet{Schutz86} to
measure directly the opacity-free $D_L$. Since the first direct
detection of the gravitational wave source GW150914
\citep{Abbott16}, as well as GW170817 \citep{Abbott17} with an
electromagnetic counterpart identified have opened an era of
gravitational wave multi-messenger astronomy \citep{Cao19}. In the
future one could use the waveform signals of GWs from inspiralling
and merging compact binaries to obtain opacity-free LD. Currently,
however, due to lack of enough GW events with identified redshifts,
extensive efforts have been focused on simulated GW data
\citep{Cai15,Cai17,Qi19a,Wei19,Wu20}. The results of these studies
demonstrated that that constraining power of GWs is comparable or
better than traditional probes, if hundreds of GW events with the
host galaxy identified are available. Recently,
\citet{Qi19a,Wei19,Liu20b} indicated that GW signals propagate in a
perfect fluid with no absorption or dissipation, which means that
information about the LDs contained in the GW signals is unaffected
by the transparency of the universe. Their work showed that using of
GW events from the third generation of GW detector, the Einstein
Telescope (ET), has significant potential and natural advantage to
test cosmic opacity. In our simulations we focus on GWs signals from
the sources with redshift up to $z \sim 5$ accessible to the
space-based GW detector-DECi-hertz Interferometer Gravitational wave
Observatory (DECIGO) which is a planed Japanese space-based GW
antenna \citep{Seto01}. The objective of DECIGO is to detect various
kinds of gravitational waves in the frequency range $0.1\sim 100$
Hz. Concerning inspiraling compact bnaries, they could be detected
by DECIGO years before they enter the sensitivity band of ground
based detectors. This will open a new window of multifrequency
gravitational wave astronomy. We will explore the potential of
DECIGO to test the transparency of universe.

For this purpose we additionally need to have independent probes of
opacity-dependent LDs extending to such high redshift range. In
particular we will consider the following probes. First is the
sample of 156 HII galaxies, which includes 25 high-$z$ HII galaxies,
107 local HII galaxies, and 24 giant extragalactic HII regions
covering the range of redshifts $0<z<2.33$. Second, we use 1048
newly-compiled SNe Ia data (Pantheon sample) covering the redshift
range of $0.07<z<2.26$. Another such probe is the sample 1598 the
quasar's having UV and X-ray flux measurements spanning the redshift
range of $0.036<z<5.10$. The purpose of our research is to assess
the precision level of opacity constraints which might be achieved
using future GW observations from the DECIGO. The paper is organized
as follows. The details of simulations of LDs accessible from the GW
data from the DECIGO are presented in Section 2. Section 3
introduces the observational data including HII galaxies and HII
regions sources, Pantheon sample, and quasars samples. Section 4
presents the cosmic opacity parameterizations used in our study and
reveals the results. Finally, we summarize our main conclusions and
make a discussion in Section 5.

\begin{figure}
\begin{center}
\includegraphics[width=0.9\linewidth]{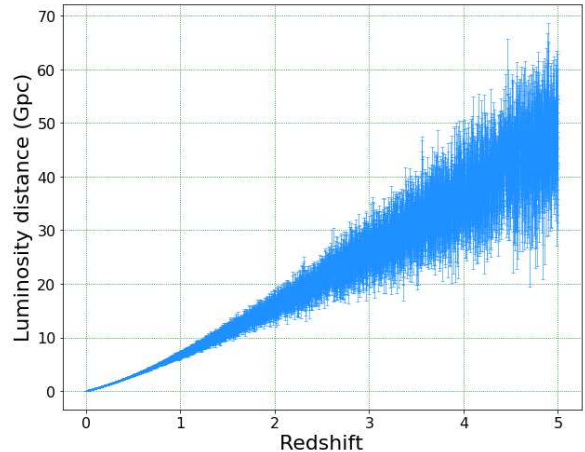}
\end{center}
\caption{ The luminosity distance measurements from 10,000
simulated GW events observable by the space detector DECIGO.}
\end{figure}

\section{GW standard sirens from DECIGO }

It is well known that GW from inspiraling binaries can be standard
sirens \citep{Abbott16,Abbott17}. More importantly, since GWs travel
in the universe without any absorption or scattering, the LDs
derived from GW signals would represent LDs unaffected by cosmic
opacity \citep{Qi19b,Zhang20}.

The DECIGO \citep{Seto01,Kawamura11}, or DECi-Hertz Interferometric
Gravitational Observatory, is Japan's proposed new mission based on
the laser interferometer space satellites. Its advantage is that
compared to LIGO, Virgo, ET and the Laser Interferometric Space
Antenna (LISA), its scientific goal lies in frequency range of
gravitational wave signal, lower than ground based detectors can
achieve, but higher that achievable for LISA configuration. Hence
the frequency gap between LISA and ground based detectors will be
filled. Moreover, GW from the inspiraling and merging compact
binaries systems consisting of neutron stars (NSs) and black holes
(BHs) can be used to determine the Hubble constant if the redshift
of the source is available \citep{Schutz86}. DECIGO will enable
discovery of NS-NS and BH-NS binaries in their inspiral phase long
time before they enter the LIGO frequency range. Therefore synergy
between DECIGO and ground based detectors will increase the
precision of inferences made from chirp signals \citep{Ola20}.
Signals in the DECIGO band can also be generated from the primordial
black holes, facilitating their ultimate detection. In addition, the
SNR ratio of DECIGO is much higher than that of current ground-based
gravitational wave detectors, so the systematic error caused by
observation will be greatly reduced, and the detection limit will be
extended to the earlier period in the history of the universe.

The Fourier transform of the GW waveform with a coalescing
binary system of masses $m_1$ and $m_2$ can be expressed as
\begin{equation}
    \widetilde{h}(f)=\dfrac{A}{D_{L}(z)}M_{z}^{5/6}f^{-7/6}e^{i\Psi(f)}\text{,}
\end{equation}
where $\Psi(f)$ is the inspiral phase term of the binary system and
$A=(\sqrt{6}\pi^{2/3})^{-1}$ is a geometrical average over the
inclination angle of a binary system. Here $D_{L}(z)$ is LD of the
source at redshift $z$ and $M_{z}=(1+z)\eta^{3/5}M_{t}$ is the
redshifted chirp mass, where $M_{t}=m_{1}+m_{2}$ is the total mass,
and $\eta=m_{1}m_{2}/M_{t}^{2}$ is the symmetric mass ratio.

The function $\Psi(f)$ represents the frequency-dependent phase
arising from the orbital evolution which can be given by 1.5 (or
higher) post-Newtonian (PN) approximation
\citep{Maggiore08,Cutler94,Kidder93}. Its concrete expression will
not affect the final result of luminosity distance calculation,
because this term will be eliminated during the detailed
calculations. Here we just need to remind that it is a function of
the coalescence time $t_{c}$, the initial phase of emission
$\phi_{c}$, $M_{z}$, $f$ and $\eta$. There are five unknown
parameters, namely: $\theta = \{ M_{z}, \eta, t_{c}, \phi_{c}, D_{L}
\}$, where $D_{L}$ is the LD at redshif $z$ determined from the rate
of changing of the frequency $f$. Following the previous analysis of
\citet{Sathyaprakash2010,Zhao11}, the mass of each neutron star is
assumed to be uniformly distributed in the range of [1,2]
$M_{\bigodot}$, while the coalescence time and the initial phase
of emission are taken as the simplified case of
$(t_{c},\phi_{c})=(0,0)$. For the purpose of uncertainty estimation
we used the Fisher matrix approach. In our case the Fisher matrix
has the following form:
\begin{equation}
 \Gamma_{ab}=4 Re \int_{f_{min}}^{f_{max}}  \dfrac{ \partial_{a} \tilde{h}^{*}_{i}(f)   \partial_{b} \tilde{h}_{i}(f)    }{S_h(f)}  df \text{,}
\end{equation}
where $\partial_{a}$ means derivative with respect to parameter
$\theta_{a}$. DECIGO design assumes the constellation of four
equilateral triangular units equipped with detectors at the vertices
of triangles. This design is equivalent to eight effective L-shaped
detectors \citep{Kawamura11,Yagi11}, so $\Gamma_{ab}$ above should
be multiplied by 8. The analytical fit of DECIGO noise spectrum
\citep{Kawamura19,Kawamura06,Nishizawa10,Yagi11} is given by
\begin{equation}
\begin{split}
S_h(f)=&6.53\times10^{-49} \left[1+(\dfrac{f}{7.36Hz})^{2} \right] \\
&+4.45\times10^{-51}\times (\dfrac{f}{1Hz})^{-4}\times \dfrac{1}{1+(\dfrac{f}{7.36Hz})^{2}}  \\
&+4.94\times10^{-52}\times (\dfrac{f}{1Hz})^{-4}  \rm{Hz^{-1}} \text{,}
\end{split}
\end{equation}
where the three terms on the right hand side represent the  shot
noise, the radiation pressure noise and the acceleration noise,
respectively.

For the purpose of the simulation, we assume the flat $\Lambda$CDM
universe as our fiducial model. The matter density parameter
$\Omega_m=0.315$ and the Hubble constant $H_0=67.4\;  km\; s^{-1}
Mpc^{-1}$ from the latest \textit{Planck} CMB observations
\citep{Planck18} are assumed in the simulations. The luminosity
distance for a flat $\Lambda$CDM universe is
\begin{equation}
D_{L,GW}(z)=\frac{c(1+z)}{H_0}\times\int^z_0\frac{dz'}
{\sqrt{\Omega_m(1+z')^3+(1-\Omega_m)}}.
\end{equation}
We put GW subscript to denote that it represents opacity-free LD of
the source determined with some uncertainty (see below). We then
make Monte Carlo simulations of $D_L(z)$ probing the distribution of
GW sources. Redshift distribution of GW sources observed on Earth
can be written as \citep{Sathyaprakash2010}
\begin{equation}
P(z)\propto\frac{4\pi D_C^2(z)R(z)}{H(z)(1+z)},
\end{equation}
where $H(z)$ and $D_C(z)$ represent the Hubble parameter and the
co-moving distance at redshift $z$, respectively. The NS-NS
coalescence rate $R(z)$ is taken the form provided by
\citet{Cutler06,Schneider01}.

Concerning the uncertainty, let us remind that the combined SNR for
the network of space borne detectors not only helps us confirm the
detection of GW with $\rho_{net}>8$, which is the SNR threshold
currently used by LIGO/Virgo network, but also contributes to the
error on the luminosity distance as
$\sigma^{inst}_{D_{L,GW}}\simeq\frac{2D_{L,GW}}{\rho}$
\citep{Zhao11}. Meanwhile, the lensing uncertainty caused by the
weak lensing should also be taken into consideration. It can be
modeled as $\sigma^{lens}_{D_{L,GW}}/D_{L,GW}=0.05z$
\citep{Sathyaprakash2010}. Therefore, the distance precision per GW
is taken as
\begin{eqnarray}
\sigma_{D_{L,GW}}&=&\sqrt{(\sigma_{D_{L,GW}}^{\rm inst})^2+(\sigma_{D_{L,GW}}^{\rm lens})^2} \nonumber\\
            &=&\sqrt{\left(\frac{2D_{L,GW}}{\rho}\right)^2+(0.05z D_{L,GW})^2}.
\label{sigmadl}
\end{eqnarray}
Now the key question is how many GW events can be detected per year
in DECIGO. According to \citet{Kawamura19} summarizing the
scientific objectives of DECIGO, 10,000 NSs binary GW signals per
year are expected to be detected at redshift $\sim$5 based on the
frequency of the NS binary coalescences given above. Even though GW
may provide us some information about the redshift of the source
\citep{Messenger12,Messenger14}, it is not enough to provide
accurate redshifts. Therefore, the electromagnetic counterparts are
necessary for the redshift determination. \citet{Cutler09}
demonstrated that it is technologically viable. Thus, we simulate
10,000 GW events and assume that their redshift is known and use
this simulation for statistical analysis in the next section. The
simulated sample of 10,000 $D_L(z)$ measurements from GW signals
detectable by the space detector DECIGO are shown in Fig.~1.

\begin{figure}
\begin{center}
\includegraphics[width=0.95\linewidth]{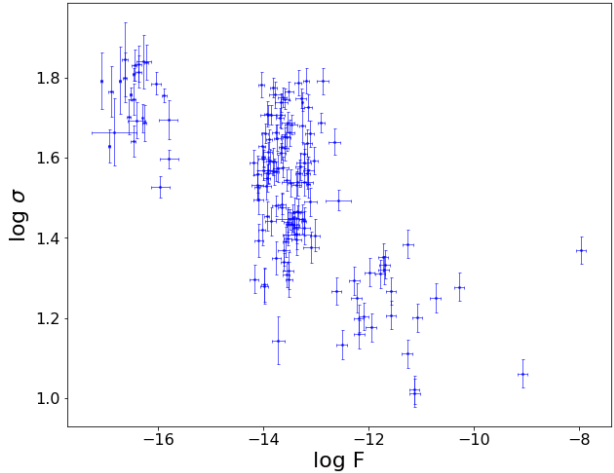}
\end{center}
\caption{The scatter plot of 156 HII galaxies and
extragalactic HII regions \citep{Terlevich2015}, with the
reddening-corrected fluxes $\log F(\textrm{H}\beta)$ and the
corrected velocity dispersions $\log \sigma(\textrm{H}\beta)$.}
\end{figure}

\begin{figure}
\begin{center}
\includegraphics[width=0.95\linewidth]{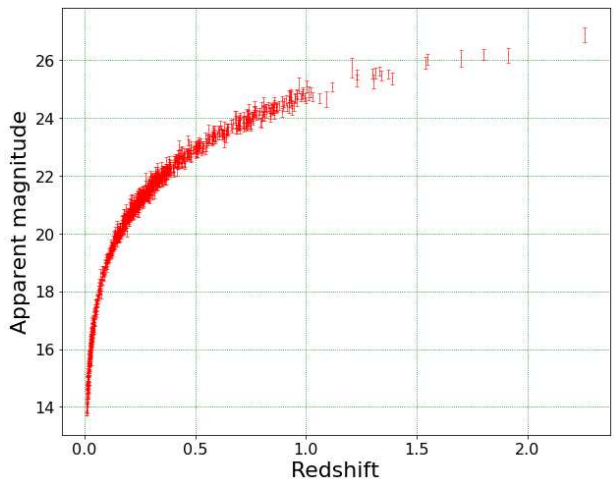}
\end{center}
\caption{The scatter plot of 1048 SNe Ia in the Pantheon
sample \citep{Scolnic18}, with the measurements of the apparent
magnitude in $B-$band ($m_B$).}
\end{figure}

\section{Opacity-dependent luminosity distances in the EM domain}

In order to get the opacity-dependent luminosity, we turn to three
catalogues of $D_L(z)$ acquired by three different methods
separately based on HII galaxies and extragalactic HII region
sources, SNe Pantheon sample, and the UV-X ray quasar sample.

\subsection{HII galaxies and extragalactic HII regions}

The first opacity-dependent LD is inferred from HII galaxies (HIIGx)
and giant extragalactic HII regions (GEHR). The existence of HIIGx
and GEHR both are explained by star formation theory. A little
different than HIIGx, GEHRs are generally located in the outer discs
of late-type galaxies, but they form physically similar systems
\citep{Melnick18}. In particular, their total luminosity is almost
completely dominated by the starburst episode and optical spectra
are indistinguishable.  The rapidly forming stars surrounded by
hydrogen ionized by massive star clusters emit strong Balmer lines
in H$\alpha$ and H$\beta$. Possibility of using these galaxies as
standard candles lies in the fact that the luminosity $L$(H$\beta)$
in H$\beta$ line is strongly correlated with the ionized gas
velocity dispersion $\sigma$. The number of energetic photons
capable of inonizing hydrogen and the turbulent velocity of the gas
both increase with the increasing mass of the starburst component
\citep{Terlevich81}.

A well-compiled sample of HII galaxies with accurately measured flux
density and the turbulent velocity of the gas could be used as a
cosmic distance indicator at redshifts beyond the current reach of
SNe Ia
\citep{Siegel2005,Plionis2011,JunJie2016,Wu20,Chavez2012,Chavez2014,Terlevich2015}.
Based on the emission-line luminosity versus ionized gas
velocity dispersion relation (called $L-\sigma$ relation), HIIGX and
GEHR used as standard candles are suitable for cosmological
applications. The relevant form of $L-\sigma$ relation reads:
\begin{equation}
\log L(\textrm{H}\beta) = \alpha \log \sigma(\textrm{H}\beta)+\kappa,
\end{equation}
where $\alpha$ is the slope and $\kappa$ is intercept. With a
well-known expression $L=F\times 4\pi D_L^2$, one can derive LD, as a function of flux $F$:
\begin{equation}
D_{L,HII} = 10^{0.5[\alpha \log \sigma(\textrm{H}\beta)-\log
F(\textrm{H}\beta)+\kappa]-25.04}.
\end{equation}
The corresponding uncertainty of $D_{L,HII}$ can be calculated by error propagation formula:
\begin{equation}
\sigma_{D_{L,HII}}=(0.5\ln10) D_{L,HII}\sqrt{(\alpha \sigma_{\log
\sigma})^{2}+(\sigma_{\log F})^{2}},
\end{equation}\label{sigmadl1}
where $\sigma_{\log \sigma}$ and $\sigma_{\log F}$ are the
observational uncertainties of the (logarithms of) reddening
corrected H$\beta$ flux and the corrected velocity dispersion.
However, it should be noted that a zero-point of the original
$L-\sigma$ relation is required to assess the model parameters,
i.e., $\alpha$, $\kappa$. Numerous efforts have been made to
calibrate the zero-point of this relation, for example, \citet{Wu20}
used Hubble parameter $H(z)$ measurements and GW to constrain model
parameters. However, in order to avoid introducing more systematic
errors in our work, we regard $\alpha$ and $\kappa$ as statistical
nuisance parameters \citep{JunJie2016}. Full information about the
sample of 156 HII regions that remain after the aforementioned
selection can be found in Table 1 of \citet{JunJie2016}. Fig.~2
shows the total sample of 156 HII source data as a function of
redshift ($0<z<2.33$), which comprises 25 high-$z$ HIIGx, 107 local
HIIGx, and 24 GEHR \citep{Terlevich2015}.

\begin{figure}
\begin{center}
\includegraphics[width=0.95\linewidth]{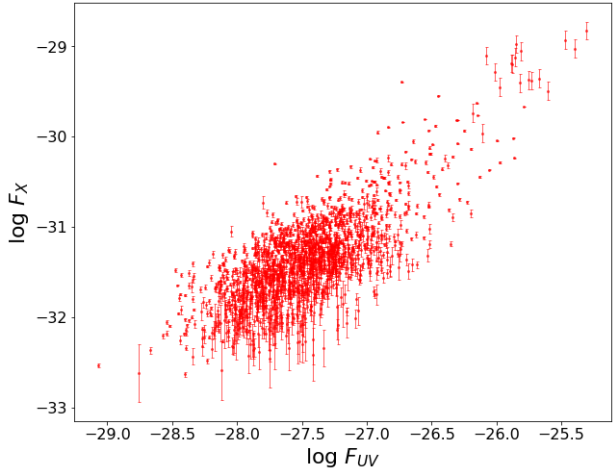}
\end{center}
\caption{The scatter plot of a sample of 1598 quasars
\citep{Risaliti2018}, with the available observations of X-ray
fluxes $\log(F_{X})$ and UV fluxes $\log(F_{UV})$.}
\end{figure}

\subsection{Type Ia supernovae observation and Pantheon dataset}

The second source of opacity-dependent LDs are SNe Ia. Recently, the
Pan-STARRS1 (PS1) Medium Deep Survey has released a dataset
(Pantheon) which contains 1048 SNe Ia covering redshift range of
$0.01<z<2.3$ \citep{Scolnic18}. The advantage of using this dataset
is richness of the sample and its depth in redshift superseding
previous datasets like JLA \citep{Betoule14}. The Pantheon catalogue
combines the subset of 279 PS1 SNe Ia \cite{Rest14,Scolnic14} and
useful distance estimates of SNe Ia from SDSS, SNLS, various low
redshift and HST samples \citep{Scolnic18}. The distance modulus of
SNe Ia can be expressed in following form
\begin{equation}
\mu_{SN}=m_B-M_B+\alpha^*\cdot X_1-\beta\cdot \mathcal{C},
\end{equation}
where $m_B$ is the apparent magnitude in $B-$band, $X_1$ is the
stretch determined by the shape of the SNe light curve,
$\mathcal{C}$ is the color measurement, $\alpha^*$ and $\beta$
coefficient characterized stretch-luminosity and color-luminosity
relationships, respectively. One should note that $M_B$ treated here
as nuisance parameter, denotes the absolute magnitude in $B-$band,
whose value is determined by the host stellar mass $M_{stellar}$ by
a step function
\begin{equation}
M_B=\left\{
   \begin{array}{ll}
   M_B^1, \,\,\,\,\,\,\,\,\,\,\,\,\,\,\,\,\,\,\,\,\,\,\,\,$if$ \,\,\,\,\,\, M_{stellar}<10^{10}M_{\odot} \\
   M_B^1+\Delta M,\,\,\,\,\,\, $otherwise$,
   \end{array}
  \right.
\end{equation}
Hence, there are four nuisance parameters $(\alpha^*, \beta,
M_B^1,\Delta M)$ to be fitted. Fortunately, the Pantheon sample is
based on the approach called BBC (for BEAMS with Bias Corrections)
\citep{Kessler17}, which is novel with respect to previous SALT2
methodology. Applying the BBC method, \citet{Scolnic18} reported the
corrected apparent magnitude $m_{B,corr}=m_B+\alpha^*\cdot
X_1-\beta\cdot \mathcal{C}+\Delta M$ for all the SNe Ia. Therefore,
the observed distance modulus of SNe Ia is simply
$\mu_{SN}=m_{B,corr}-M_B$ \citep{Ma19}. The total uncertainty of
distance modulus in the Pantheon dataset can be expressed
\begin{equation}
\sigma_{\mu_{SN}}=\sqrt{\sigma_{m_{B,corr}}^2},
\end{equation}
where $\sigma_{m_{B,corr}}$ is the observational error of the
corrected apparent magnitude (see \citet{Scolnic18} for more
details). One can get the LD of SNe Ia in Mpc by a well-known
equation
\begin{equation}
D_{L,SN}=10^{\mu_{SN}/5-5},
\end{equation}
and the corresponding uncertainty is
\begin{equation}
\sigma_{D_{L,SN}}=(\ln10/5) D_L\sigma_{\mu_{SN}}.
\end{equation}\label{sigmadl2}
Fig.~3 shows the SNe Ia data from the most recently compiled
Pantheon sample \citep{Scolnic18}, as a function of redshift ranging
from 0.01 to 2.3.

\begin{figure}
\begin{center}
\includegraphics[width=0.95\linewidth]{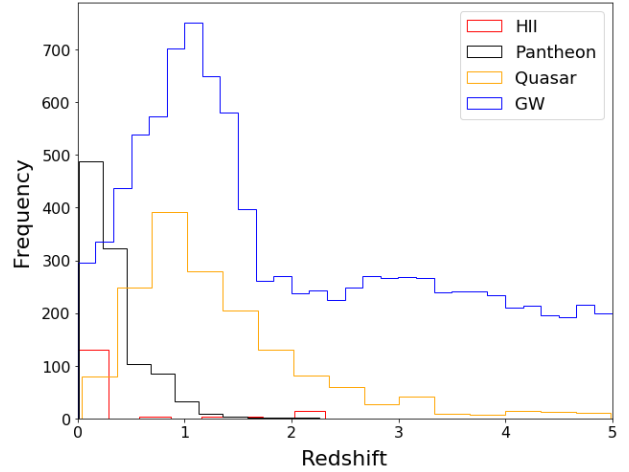}
\end{center}
\caption{ The redshift distribution of NS-NS binaries that
would be observed by DECIGO, compared with distribution of EM probes
used to derive opacity-dependent luminosity distance.} 
\end{figure}

\subsection{Nonlinear relation between quasar's X-ray and UV flux measurements}

Being the brightest sources in the universe, quasars (QSO) have
considerable potential for being useful cosmological probes
\citep{Ma17,Xu18,Cao18,Cao19b,Cao20,Zheng20}. Unlike SNe Ia, quasars
display extreme variability in luminosity and high observed
dispersion, which prevents quasars for being a standard candle with
a high precision. However, \citet{Risaliti2015} through the analysis
and refinement of the QSO sample with well measured X-ray and UV
fluxes eliminated considerably the contribution of observational
dispersion paving the way to make them standard candles .
Subsequently, \citet{Risaliti2018} collected a sample of 7238
quasars with available X-ray and UV measurements from the
cross-correlation of the XMM-Newton Serendipitous Source Catalogue
Data Release 7 \citep{Rosen2016} and the Sloan Digital Sky Survey
(SDSS) quasar catalogues from Data Release 7 \citep{Shen11} and DR12
\citep{Paris17}, and then selected 1598 high-quality QSO sample
suitable for cosmological applications. They final results showed
that a sample of 1598 quasars with available UV and X-ray
observations could produce a Hubble diagram in excellent agreement
with that of SNe Ia in the redshift range of $(z<1.4)$
\citep{Liu20c}. They used a simple non-linear relation between the
X-ray and UV luminosities \citep{Avni1986} where X-rays are Compton
scattered photons from an overlying, hot corona and the UV photons
are emitted by an accretion disk
\begin{equation}\label{xuvrelation}
\log(L_X)=\gamma\log(L_{UV})+\beta',
\end{equation}
where $L_X$ and $L_{UV}$ respectively represent the monochromatic
luminosity at 2keV and 2500${\AA}$ in rest-frame, while $\beta'$ and
$\gamma$ denote the intercept and the slope parameter. With the
combination of Eq.~(14) and a well-known equation $L=F\times 4\pi
D_L^2$, one can derive a LD for each quasar, as a function of the
fluxes $F$, the slope $\gamma$, the normalization $\beta$
\begin{equation}
D_{L,QSO}=10^{\frac{1}{2-2\gamma}\times[\gamma\log(F_{UV}) - \log(F_X) + \beta]},
\end{equation}
where $\beta$ is a constant that subsumes the slope $\gamma$ and the
intercept $\beta'$, such that $\beta=\beta'+(\gamma-1)\log 4\pi$.
From theoretical point of view, LD can be directly determined from
the measurements of the fluxes of $F_X$ and $F_{UV}$, with a
reliable knowledge of the value for the two parameters ($\gamma$,
$\beta$) characterizing the $L_X - L_{UV}$ relation as well as its
dispersion. Because the measurement uncertainty of $F_{UV}$ is
smaller than the uncertainty of $F_{X}$ and the global intrinsic
dispersion, we therefore neglect it. Thus, the uncertainty of LD
from QSO sample is given by
\begin{equation}
\sigma_{D_{L,QSO}}=\ln 10 / (2-2\gamma) D_{L,QSO}\sqrt{\sigma_{\log(F_X)}^2+\delta^2}.
\end{equation}\label{sigmadl3}
The scatter plot of 1598 quasar sample reported by
\citet{Risaliti2018} is shown in Fig.~4, with the redshift ranging
from 0.036 to 5.1.

It should be recalled that the cosmic absorption could affect LD
measurements from all the above mentioned probes accessible in the
EM window. LD measurements from GWs being unaffected by cosmic
opacity, are related to $D_{L,obs}$ representing the LDs derived
from HII, SNe, or QSO according to
\begin{equation}
D_{L,obs}(z)=D_{L,GW}(z)e^{\tau(z)/2}.
\end{equation}
where $\tau(z)$ is the optical depth related to the cosmic
absorption. More specifically, if the universe is opaque, the number
of photons from sources received by the observer will be reduced,
i.e., if $\tau>0$ which means the reduction of the number of
photons.

\section{Methodology and constrained results of cosmic opacity}

Let us note that all three kinds of probes used to derive
opacity-dependent LDs contain model parameters. One possible way to
determine their values could be by using external calibrators.
However, this might introduce extra systematics and biases, which
are hard to quantify. Therefore, in our work, we regard them as free
parameters to be optimized along with cosmic opacity parameter
$\epsilon$ (definition is given below). To be more specific, they
are: ($\alpha$, $\kappa$) in the case of HII galaxies and regions,
($M_B$) in SNe Ia Pantheon sample and ($\beta$, $\gamma$, $\delta$)
in the X-ray - UV quasar sample. Even though this procedure affects
the precision of constraints, yet it is a coherent procedure. In
order to avoid any bias coming from the redshift differences between
opacity-independent and opacity-dependent LDs, we apply the
following criterion $|z_{GW}-z_{obs}|<0.005$, where $obs$ represents
HII, SNe, or QSO. This redshift selection criterion leads to
decrease of the sample sizes. As a result 45 data point are kept for
HII regions, 897 for SNe and 1595 for quasars. The redshift
distribution of NS-NS binaries that would be observed by DECIGO is
shown in Fig.~5, together with the distribution of three types of EM
probes. One can see that current QSO sample has a very similar
distribution to future GW data obtainable by DECIGO.

It should be pointed out that, locally transparent universe (at
$z=0$) might be significantly opaque at high redshifts. Such
tendency is strongly supported by the observed rate of the damped
Ly$\alpha$ absorbers in high-redshift quasars
\citep{Prochaska04,Rao06}, as well as the reported abundance of
dusty galaxies and dusty halos around star-forming galaxies at
$z\sim 7$ \citep{Watson15,Fujimoto19}. Therefore, the cosmic opacity
-- an integral quantity dependent on the proper dust density,
grain-size distribution, and the dust extinction efficiency -- could
significantly increase with redshifts. In our analysis, such issue
is approached in two ways. Firstly, in analogy to previous works of
\citet{Li13,Liao13}, we assume two particular parameterized forms of
the optical depth $\tau(z)$:
\begin{eqnarray}
P1: \  \tau(z)& = & 2\epsilon z, \nonumber\\ P2: \   \tau(z)&= &
(1+z)^{2\epsilon}-1.
\end{eqnarray}
If $\epsilon=0$, it means that photon number is conserved and the
universe is transparent. However, one should also be aware of the
disadvantage of the first method, i.e., the opacity test just proves
whether the opacity follows a specific phenomenological function, as
stressed in \citep{Vavrycuk20}.  Therefore, in the subsequent
analysis the optical depth will also be studied in individual
redshift bins, in order to perform the opacity test in a more
generalized way and show how decisive conclusions about the
transparency of the universe can be deduced from such tests.

In order to constrain the nuisance parameters and the cosmic opacity
parameter simultaneously, we use Python Markov Chain Monte Carlo
(MCMC) module, emcee \citep{Foreman13}, to obtain the best fitted
values and corresponding uncertainties by minimizing the $\chi^2$
objective function defined as
\begin{equation}
\chi^2 = \sum_{1}^{i} \frac{[D_{L,GW}(z_i;\epsilon)-D_{L,obs}(z_i;\textbf{p})]^2}{\sigma_{D_L}(z_i)^2},
\end{equation}
where the $\textbf{p}$ collectively denote nuisance parameters of
respective model,
$\sigma_{D_L}^2=\sigma_{D_{L,GW}}^2+\sigma_{D_{L,obs}}^2$,
$\sigma_{D_{L,GW}}$ and $\sigma_{D_{L,obs}}$ are given by Eq.~(5)
and Eqs.~(8), (13), (16), respectively.

\begin{figure}
\centering
\includegraphics[scale=0.6]{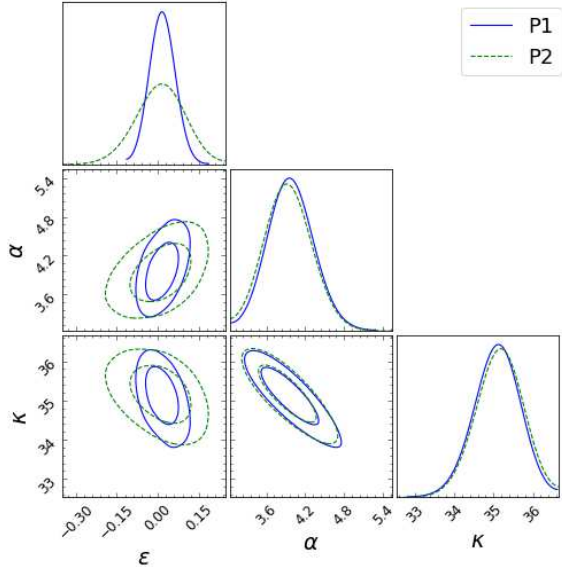}
\caption{One-dimensional distributions and two-dimensional
constraint contours for cosmic opacity parameter $\epsilon$ and HII
regions nuisance parameters ($\alpha$, $\kappa$) in both P1 (blue
dashed line) and P2 parametrization (green solid line).}
\end{figure}

\begin{figure*}
\centering
\includegraphics[scale=0.55]{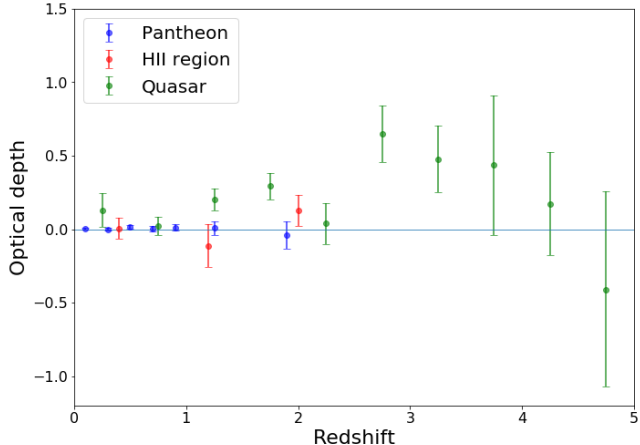}
\caption{ Optical depth calculated from the combinations of
available EM observations and the GW sample observed by DECIGO: the
binned optical depth with the 68.3\% confidence intervals
corresponding to the redshift bins of [0, 0.8, 1.6, 2.4] for HII
region sample, the redshift bins of [0, 0.2, 0.4, 0.6, 0.8, 1.0,
1.5, 2.3] for the SNe Ia Pantheon sample, and the redshift bins of
[0, 0.5, 1.0, 1.5, 2.0, 2.5, 3.0, 3.5, 4.0, 4.5, 5.0] for the quasar
sample.}
\end{figure*}

\begin{table*}
\begin{center}
\begin{tabular}{c| c c c c c c c}
\hline
Parameterized form+Sample   & $\epsilon$ & $\alpha$ &$\kappa$ &$M_B$ &$\beta$ &$\gamma$ &$\delta$\\
\hline
\hline
P1+HII    & $0.016^{+0.037}_{-0.036}$ & $3.966^{+0.283}_{-0.261}$ & $35.112^{+0.426}_{-0.465}$& $\Box$ & $\Box$ & $\Box$ & $\Box$ \\
\hline
P2+HII    & $0.013^{+0.063}_{-0.071}$ & $3.924^{+0.278}_{-0.254}$ & $35.175^{+0.416}_{-0.453}$& $\Box$ & $\Box$ & $\Box$ & $\Box$ \\
\hline
P1+SN   & $0.006^{+0.009}_{-0.009}$& $\Box$ & $\Box$ & $-19.414^{+0.008}_{-0.008}$ & $\Box$ & $\Box$ & $\Box$ \\
\hline
P2+SN   & $0.009^{+0.012}_{-0.013}$& $\Box$ & $\Box$ & $-19.415^{+0.009}_{-0.008}$ & $\Box$ & $\Box$ & $\Box$ \\
\hline
P1+QSO  & $0.056^{+0.032}_{-0.036}$& $\Box$ & $\Box$ & $\Box$ & $7.550^{+0.408}_{-0.408}$ & $0.623^{+0.014}_{-0.014}$ & $0.231^{+0.004}_{-0.004}$  \\
\hline
P2+QSO  & $0.110^{+0.058}_{-0.080}$& $\Box$ & $\Box$ & $\Box$ & $7.526^{+0.423}_{-0.423}$ & $0.624^{+0.015}_{-0.014}$ & $0.231^{+0.004}_{-0.004}$  \\
\hline
\hline
\end{tabular}
\caption{Results from three types of GW+EM data combinations: the
best-fitted value and $68\%$ C.L. of $\epsilon$ and other nuisance
parameters in two cosmic-opacity parameterizations.}
\end{center}
\end{table*}

In Fig.~6, we plot the one-dimensional marginalized distributions
and two-dimensional constraint contours for the opacity parameter
$\epsilon$ with HII regions nuisance parameters ($\alpha$, $\kappa$)
in the P1 and P2 parametrizations. In the P1 case, we obtain
$\epsilon=0.016^{+0.037}_{-0.036}$ with the model nuisance
parameters $\alpha=3.966^{+0.283}_{-0.261}$ and
$\kappa=35.112^{+0.426}_{-0.465}$ at 68.3\% confidence level. The
1$\sigma$ confidence region constraint on cosmic opacity $\epsilon$
in P2 case is $\epsilon=0.013^{+0.063}_{-0.071}$. Our results show
that simulated GW data tested against HII regions reaching to
redshift $z<2.316$, support the transparent universe in each
parameterized function of optical depth $\tau(z)$. Using different
parametrizations of optical depth yields slightly different
constraints of the cosmic opacity $\epsilon$ and model nuisance
parameters. In the P1 case GW data from the DECIGO would be able to
constrain cosmic transparency with the precision of $\Delta
\epsilon\sim0.04$. It is worth remembering that due to our redshift
pre-selection a sample of only 45 HII regions were used here to
combine with GW events from the DECIGO. More interestingly, we find
that our constraints on the nuisance parameters (see Table I) are
different from those results of \cite{Wu20}: $\alpha=5.17\pm0.09$,
$\kappa=32.86\pm0.12$, which are derived from a calibration by
simulated GW events in a flat $\Lambda$CDM cosmology. The main
reason for the difference is that only 45 HII regions samples are
used here rather than 156 sources in work of \cite{Wu20}. The other
reason may be due to the covariance between cosmic opacity and the
nuisance of HII regions, which can be seen from the constraint
contours.

One should bear in mind that, an accurate reconstruction of
$\tau(z)$ can considerably improve our understanding of the nature
of cosmic opacity. In order to reconstruct the evolution of
$\tau(z)$ without any prior assumption of its specific functional
form, we calculate the optical depth from Eq.~(18) at individual
redshifts and average $\tau(z)$ within redshift bins. Fig.~7 shows
the resulting optical depth. One can see that no statistically
significant evidence of deviation from the transparent universe
($\tau(z)=0$) can be detected from low-redshift HII galaxies. Yet,
interestingly, the HII region data favors a transition from $\tau=0$
at low redshift to $\tau>0$ at higher redshift, a behavior that is
consistent with the true optical depth increasing with redshifts.
Such tendency, which is different from that obtained in the
framework of parameterized form, is supported by the recent works of
\citet{Ma18,Vavrycuk20} stressing the importance of cosmic-opacity
test without a prescribed phenomenological function.

\begin{figure}
\centering
\includegraphics[scale=0.55]{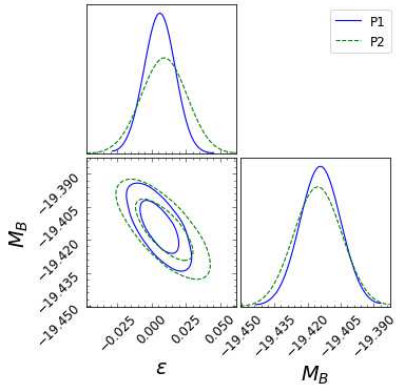}
\caption{One-dimensional distributions and two-dimensional
constraint contours for cosmic opacity parameter $\epsilon$ and SNe
Ia nuisance parameters ($M_B$), using the Pantheon sample and
simulated GW dataset.}
\end{figure}

The results derived from the simulated GW events from DECIGO and
Pantheon sample are shown in Fig.~8 and Table I. One can see that
almost transparent universe is also favored up to  redshift
$z<2.26$: the best fitted value of $\epsilon$ with 1$\sigma$ C.L. is
$0.006\pm{0.009}$ and $0.009\pm{0.012}$ for P1 and P2 cases,
respectively. Compared with previous work, \citet{Qi19a} combined
simulated GW events from the ET with Pantheon SNe Ia to test the
opacity of the Universe, they results shown that the best-fit values
with 1$\sigma$ standard error for cosmic opacity is
$0.009\pm{0.016}$ with the absolute magnitude $M_B=-19.4 \pm0.016$
in P1 case ($\epsilon=0.015\pm 0.025$ and $ M_B=-19.404 \pm0.019$ in
P2 case). Although their results are consistent with our work (not
only $\epsilon$, but also $M_B$), our finding suggest that the
future space-based GW detector DECIGO could result in more stringent
constraints on the transparency of the universe than the ET.
Meanwhile, the optical depth $\tau(z)$ is also calculated by
combining the simulated GW events and the current SNe Ia
observations at individual redshift bins, with the final results
displayed in Fig.~7. Benefit from the large sample size of the
Pantheon SNe Ia sample, the cosmic opacity can be constrained with
higher precision. As can be seen from individual points of $\tau(z)$
as well as for its binned values, the optical depth without any
parameterized form show that the universe is transparent within
1$\sigma$ C.L. However, we also find that a redshift bin shifts the
confidence interval for $\tau(z=0.5)$ towards large optical depth,
which might be overlooked by the cosmic-opacity test in a prescribed
phenomenological function.

The use of the QSO sample extending up to $z\sim 5$, in combination
with GW data supposed to be obtainable in DECIGO, enables us to
probe the transparency of the universe at much earlier stages of its
evolution. Fig.~9 displays the constraint results on cosmic opacity
$\epsilon$ with quasar nuisance parameters ($\beta$, $\gamma$,
$\delta$) in the P1 and P2 cases. For the first P1 parametrization
form, we obtain $\epsilon=0.056^{+0.032}_{-0.036}$,
$\beta=7.550^{+0.408}_{-0.408}$, $\gamma=0.623^{+0.014}_{-0.014}$,
$\delta=0.231^{+0.004}_{-0.004}$ where upper and lower number denote
$68\%$ confidence region. For the second P2 parametrization form,
the results are $\epsilon=0.110^{+0.058}_{-0.080}$,
$\beta=7.526^{+0.423}_{-0.423}$, $\gamma=0.624^{+0.015}_{-0.014}$,
$\delta=0.231^{+0.004}_{-0.004}$. Our results indicate that current
observations of QSOs confronted with simulated opacity-free LDs from
GW signals do not suggest any significant deviation from the
transparency of the universe. A transparent universe is supported by
the future GW dataset simulated from the DECIGO and quasars flux
measurements within 2$\sigma$ confidence level. One can clearly see
that using quasars sample with the different parameterized forms for
optical depth result with significantly different central values of
$\epsilon$ than HII regions or the Pantheon sample. \citet{Qi19a}
suggested that the dependence of test cosmic opacity on the optical
depth parameterization chosen for is relatively weak. This was
because the redshift range explored in their work was only up to
$z\sim 2.3$, while in our case it is $z\sim 5$. The two
parameterized forms P1 and P2 are approximately equal at low
redshifts, but at high redshifts they diverge. Our results
highlights the importance of choosing a reliable parameterization
for $\tau(z)$ in order to better check the cosmic opacity validity
at much high redshift. Although our best fitted central value of
cosmic opacity parameter has some deviation from $\epsilon=0$, yet
the quasars' nuisance parameters are consistent at 1$\sigma$
confidence level with the results of \citet{Melia19}, which were
obtained within the flat $\Lambda$CDM cosmology. More importantly,
our findings show a strong degeneracy between the opacity parameter
$\epsilon$ and quasars nuisance parameters $\beta$ and $\gamma$. A
little change in $\beta$ or $\gamma$, will change the value of
opacity parameter $\epsilon$ noticeably. This covariance is similar
to that seen in the SNe Ia. Therefore, using an independent reliable
external calibrator to calibrate quasar's nuisance parameter becomes
particularly important. In order to make the opacity test to more
general, optical depth is also studied in individual redshift bins
by combining quasar sample and simulated GW events from DECIGO. The
results are shown in Fig.~7. Compared with the results obtained from
the HII galaxies and Pantheon SNe Ia, there is a noticeable
improvement in precision when the quasar sample is considered. More
importantly, we obtain that non-zero optical depth is statistically
significant only for redshift bins $0<z<0.5$, $1<z<2$, and
$2.5<z<3.5$. More data extending to the above redshifts will be
necessary to investigate the cosmic opacity in this high-redshift
region where the uncertainty is still very large.

\begin{figure}
\centering
\includegraphics[scale=0.38]{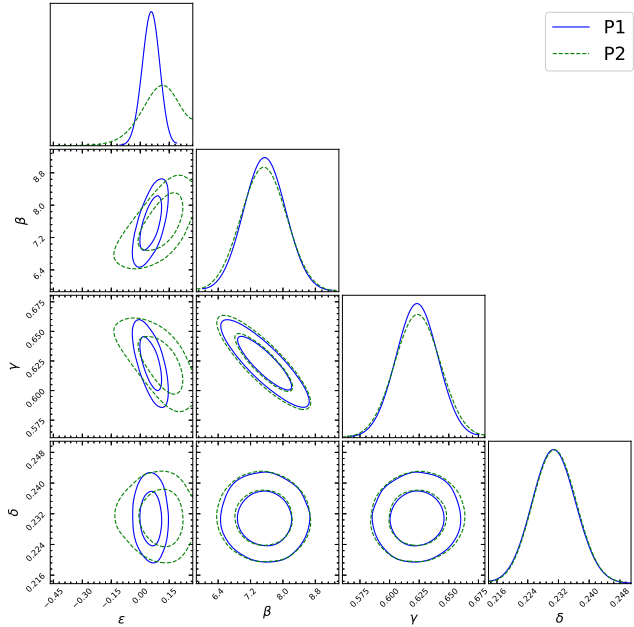}
\caption{One-dimensional distributions and two-dimensional
constraint contours for cosmic opacity parameter $\epsilon$ and
quasar nuisance parameters ($\beta$, $\gamma$, $\delta$), using the
QSO sample and simulated GW dataset.}
\end{figure}

In order to highlight the advantages of our work, it is worthwhile
to compare our results with the previous analysis performed to test
the cosmic opacity with actual or expected tests involving the LDs
from various astrophysical probes. In the work of \citet{Wei19}, the
authors combined the GW observations of a third-generation GW
detector -- ET with SNe Ia data Pantheon sample in similar redshift
ranges to test cosmic opacity. Their results shown that the cosmic
opacity parameter $\Delta \epsilon \sim0.026$ (at 68.3\% confidence
level) for P1 function with 1000 simulated GW events, and they
concluded that using GW standard sirens and SNe Ia standard candles
yields the results competitive with previous works. Moreover,
\citet{Qi19a} performed the same test of cosmic opacity as
\citet{Wei19} did. Differently, \citet{Qi19a} considered not only
the Pantheon sample but also the JLA sample. Combing luminosity
distances from joint light-curve analysis (JLA) and simulated GW
events from ET, they got $\epsilon=0.002\pm0.035$ for P1
parametrization and $\epsilon=-0.0060\pm0.053$ for P2 case,
respectively. It is necessary to mention that although we expect ten
thousand GW signals from DECIGO detectable, yet after matching with
the Pantheon sample, only about 1000 GW signals satisfy the redshift
selection criteria, which is about the same number as in the works
of \citet{Wei19,Qi19a}. By comparing the results at 1$\sigma$
confidence level, we obtain the uncertainty smaller by 65\% than
these of \citet{Wei19} in the P1 case with Pantheon dataset.
Comparing to \citet{Qi19a}, uncertainties of our results are 75\%
and 78\% smaller than their in the P1 and P2 case, respectively.
Such stringent constraints are a result of the DECIGO data having
much higher  SNR than the ET, hence the uncertainties of the LDs
inferred are smaller. Therefore, using GW signals from the DECIGO
may achieve higher precision of the measurements of cosmic opacity
than the other popular astrophysical probes.

\section{Conclusions and discussion}

In this paper, we proposed a new model-independent cosmological test
for the cosmic opacity at redshifts up $z\sim 5$. We combined
opacity-free LDs inferred from simulated GW events representative to
the space-based GW detectors of DECIGO with three popular
astrophysical probes used to derive the  opacity-dependent LDs. The
GW signals propagate in a perfect fluid without any absorption or
dissipation, which means that the information about LDs contained in
the GW signals is  unaffected by the transparency of the universe.
Such high redshift range has never been explored by previous works.
We adopted two parameterization forms to describe the optical depth
$\tau(z)$ describing the cosmic absorption.

Different from the previous works, we not only used SNe Ia as
standard candles to obtain opacity-dependent LD, but also HII
regions sources based on $L$--$\sigma$" relation and quasar flux
measurements based on nonlinear relation in X-ray and UV bands
acting standard candles. Considering the calibration of nuisance
parameters occurring typically in standard candles, we treated them
as free parameters optimized along with cosmic opacity parameter.
One of the purposes of our research is to what extent can one test
the transparency for our universe if DECIGO detects a certain number
of GW events.

In the case of HII regions serving as standard candles, the redshift
selection resulted with only 45 HII regions data points left and
used to constrain cosmic opacity. The cosmic opacity parameter
$\epsilon$ was constrained with the precision is 0.04 and 0.07 in P1
and P2 cases, respectively. Compared with previous papers, such as
\citet{Li13}, who used two galaxy cluster samples and the Union2.1
to constrain cosmic opacity resulting with $\epsilon=0.009\pm0.057$
for P1 case and $\epsilon=0.014\pm0.070$ for P2, our results shown
that HII regions have powerful potential to get more stringent
constraint on cosmic opacity. On the other hand, we shown that just
45 GW events from the DECIGO can constrain the cosmic opacity
parameter with high accuracy. More interestingly, our constraints on
the HII regions nuisance parameters ($\alpha$, $\kappa$) are
somewhat  different from recent results of \citet{Wu20}. In the case
of SNe Ia Pantheon sample, the cosmic opacity can be constrained
with higher precision: the uncertainty can be further reduced to
$\Delta \epsilon\sim0.010$, and the absolute magnitude uncertainty
can be reduced to $\Delta \epsilon\sim0.016$, which is consistent
with the previous works \citep{Wei19,Qi19a}. By comparing our
results with previous opacity constraints, we prove that our method
using GW signals from space-based DECIGO detector as standard sirens
and SNe Ia standard candles will be competitive.

The third cosmological probe we used was a carefully selected sample
of QSOs whose X-ray and UV fluxes are measured and which are
promising new class of standard candles. Combining QSOs with
simulated GW events, our result show that, there is no large
deviation from a transparency of the Universe within $2\sigma$
confidence level. The cosmic opacity parameter was  constrained with
an accuracy $\Delta \epsilon\sim 10^{-2}$.  Due to the lack of
convincing form of parameterization for $\tau$ at such high
redshift, or any reliable theoretical suggestion for it, two
parameterizations \citep{Li13,Liao13} were used in our work. We
demonstrated that the cosmic opacity results were slightly sensitive
to the parametrization form assumed. This emphasizes the importance
of choosing a reliable parameterization form for $\tau(z)$ at much
high redshift. Moreover, our findings illustrate that there is a
strong degeneracy between the cosmic opacity parameter $\epsilon$
and quasars nuisance parameters ($\beta$, $\gamma$) which may affect
the real the value of the cosmic opacity parameter. The independent
external calibration of quasars nuisance parameters is therefore
particularly important. In order to reconstruct the evolution of
$\tau(z)$ without assuming any specific functional form of it, the
cosmic opacity tests should be applied to individual redshift bins
independently. Therefore, in this paper we calculated the optical
depth at individual redshifts and averaged $\tau(z)$ within redshift
bins. Our findings indicated that, compared with results obtained
from the HII galaxies and Pantheon SNe Ia, there is an improvement
in precision when the quasar sample is considered, while non-zero
optical depth is statistically significant only for redshifts
$0<z<0.5$, $1<z<2$, and $2.5<z<3.5$.

It should be pointed out that according to the recent analysis of
\citet{Vavrycuk19,Vavrycuk20}, the cosmic opacity could vary with
frequency since it strongly depends on wavelength according to the
extinction law \citep{Mathis90,Li01,Draine03}. The problem was
recognized a long time ago \citep{Aguirre03,Corasaniti06}, with a
heuristic suggestion that the luminosity distance data sets covering
different frequencies might be differently sensitive to the
transparency of the universe \citep{Vavrycuk20}. In our analysis, we
combined the predicted future GW observations from DECIGO with three
types of EM data (HII regions, SNe Ia Pantheon sample, quasar
sample), and furthermore checked whether these EM data can be
affected by cosmic opacity in different ways. Considering the fact
that X-ray photons can destroy dust grains instead of being
absorbed, physically different processes could be detected from the
observations of X-ray fluxes, instead of simple luminosity dimming
due to dust absorption of low-energy photons
\citep{Draine02,Morgan14}. Our final results showed that the cosmic
opacity could attain different values for different types of $D_L$
data covering different wavelengths (optical for HII regions, SNe Ia
Pantheon sample, ultraviolet for quasar sample), although such
effect is still difficult to be precisely quantified (see
\citet{Vavrycuk20} for a detailed discussion). Thus, the goal of our
test is not just to check the resolution power of the GW data, but
also to assess the performance of the combination of current and
future available data in gravitational wave (GW) and electromagnetic
(EM) domain. Let us also comment on the issue raised by
\citet{Vavrycuk20}, that combining the GW data with other types of
data (SNe Ia, QSO, HII galaxies) should be avoided since these data
can be affected by cosmic opacity in a different way, because the
cosmic opacity is strongly frequency dependent. While this point of
view is in principle true, one can argue in the following way: the
data sets of cosmological probes like standard candles have been
prepared according to the best available knowledge concerning
possible opacity sources. In particular, \citet{Risaliti2018} have
considered the (differential) absorption in UV and X-ray bands while
constructing their sample. Then, after checking against opacity-free
probes like standard rulers or GW standard sirens, by using the DDR
relation (the fundamental one, but frequency independent) one could
able to find the evidence for unaccounted opacity sources, or at the
very least the new physics involved. This is the idea behind our
approach.

As a final remark, we have successfully constrained the cosmic
opacity covering the redshift range $0<z<5$. Although the constraint
on cosmic opacity by using quasars X-ray and UV flux measurements
did not improve the results obtained with  HII regions or SNe Ia, it
provided an insight into cosmic opacity at an earlier stage of the
universe. Our results demonstrate the huge potential of standard
candles and GW signals -- standard sirens to measure cosmic opacity
at high redshifts. Current GW detectors did not allowed us yet to
make such inferences, however in the near future the data we were
forced to simulate will become a daily experience of now emerging GW
astronomy. It would be fruitful to prepare for this era and suggest
which problems could be addressed then with high accuracy. Cosmic
opacity issue certainly belongs to this class of problems.

\section*{Acknowledgments}

We would like to thank the referee for constructive
comments, which allowed to improve the paper substantially. This
work was supported by National Key R\&D Program of China No.
2017YFA0402600; the National Natural Science Foundation of China
under Grants Nos. 12021003, 11690023, 11633001, and 11373014;
Beijing Talents Fund of Organization Department of Beijing Municipal
Committee of the CPC; the Strategic Priority Research Program of the
Chinese Academy of Sciences, Grant No. XDB23000000; the
Interdiscipline Research Funds of Beijing Normal University; and the
Opening Project of Key Laboratory of Computational Astrophysics,
National Astronomical Observatories, Chinese Academy of Sciences.
J.-Z. Qi was supported by China Postdoctoral Science Foundation
under Grant No. 2017M620661, and the Fundamental Research Funds for
the Central Universities N180503014. M.B. was supported by the
Foreign Talent Introducing Project and Special Fund Support of
Foreign Knowledge Introducing Project in China. He was supported by
the Key Foreign Expert Program for the Central Universities No.
X2018002.



\begin{thebibliography}{}


\bibitem[Abbott et al.(2016)]{Abbott16} Abbott, B. P., et al. [LIGO Scientific and Virgo Collaborations], 2016, PRL, 116, 061102
\bibitem[Abbott et al.(2017)]{Abbott17} Abbott, B. P., et al. [LIGO Scientific Collaboration, the Virgo Collaboration], 2017, PRL, 119, 161101
\bibitem[Aguirre(2000)] {Aguirre03} Aguirre, A. N. 2000, ApJ, 533, 1
\bibitem[Avgoustidis et al.(2010)]{Avgoustidis10} Avgoustidis, A., Burrage, C., Redondo, J., Verde L., \& Jimenez, R. 2010, JCAP, 10, 024
\bibitem[Avni \& Tananbaum(1986)]{Avni1986} Avni, Y. \& Tananbaum, H. 1986, ApJ, 305, 83
\bibitem[Betoule et al.(2014)]{Betoule14}Betoule, M., et al. 2014, A\&A, 568, 22
\bibitem[Caldwell et al.(1998)]{Caldwell98}  Caldwell, R., et al. 1998, PRL, 80, 1582
\bibitem[Cao \& Liang(2011)]{Cao11} Cao, S., \& Liang, N. 2011, RAA, 11, 1199
\bibitem[Cao, et al.(2011)]{Cao11a} Cao, S., Liang, N., \& Zhu, Z.-H. 2011, MNRAS, 416, 1099
\bibitem[Cao \& Liang(2013)]{Cao13a} Cao, S., \& Liang, N. 2013, IJMPD, 22, 1350082
\bibitem[Cao \& Zhu(2014)]{Cao14} Cao, S., \& Zhu, Z.-H. 2014, PRD, 90, 083006
\bibitem[Cao et al.(2017a)]{Cao17a} Cao, S., Zheng X., Biesiada M., Qi J., Chen Y. \& Zhu Z.-H. 2017a, A\&A, 606, A15
\bibitem[Cao et al.(2017b)]{Cao17b} Cao, S., Biesiada, M., Jackson, J., Zheng, X. \& Zhu Z.-H. 2017b, JCAP, 02, 012
\bibitem[Cao et al.(2018)]{Cao18} Cao, S., et al. 2018, EPJC, 78, 749
\bibitem[Cao et al.(2019a)]{Cao19} Cao, S., et al. 2019a, NatSR, 9, 11608
\bibitem[Cao et al.(2019b)]{Cao19b} Cao, S., et al. 2019b, PDU, 24, 100274
\bibitem[Cao et al.(2020)]{Cao20} Cao, S., et al. 2020, ApJL, 888, L25
\bibitem[Cai et al.(2015)]{Cai15} Cai, R.-G., et al. 2016, PRD, 93, 043517
\bibitem[Cai \& Yang(2017)]{Cai17} Cai, R.-G. \& Yang, T. 2017, PRD, 95, 044024
\bibitem[Ch\'{a}vez et al.(2012)]{Chavez2012}Ch\'{a}vez, R., et al. 2012, MNRAS, 425, L56
\bibitem[Ch\'{a}vez et al.(2014)]{Chavez2014}Ch\'{a}vez, R., et al. 2014, MNRAS, 442, 3565
\bibitem[Chen(1995)]{Chen95} Chen, P. 1995, PRL, 74, 634
\bibitem[Chen(2012)]{Chen12} Chen, J., Wu, P., Yu, H., \& Li, Z. 2012, JCAP, 10, 029
\bibitem[Corasaniti(2006)] {Corasaniti06} Corasaniti, P. S. 2006, MNRAS, 372, 191
\bibitem[Csaki et al.(2002)]{Csaki02} Csaki, C., et al. 2002, PRL, 88, 161302
\bibitem[Cutler \&  Flanagan(1994)]{Cutler94} Cutler, C. \& Flanagan, E. E. 1994, PRD, 49, 2658
\bibitem[Cutler \& Harms(2006)]{Cutler06} Cutler, C. \& Harms, J. 2006, PRD, 73, 042001
\bibitem[Cutler \& Holz(2009)]{Cutler09} Cutler, C. \& Holz, D. E. 2009, PRD 80, 104009
\bibitem[Deffayet \& Uzan(2000)]{Deffayet00}Deffayet, C. \& Uzan, J.-P. 2000, PRD, 62, 063507
\bibitem[Draine \& Hao(2002)] {Draine02} Draine, B. T. \& Hao, L. 2002, ApJ, 569, 780
\bibitem[Draine(2003)] {Draine03} Draine, B. T. 2003, ARA\&A, 41, 241
\bibitem[Ellis(2007)]{Ellis}  Ellis, G. F. R. 2007, Gen. Rel. Grav., 39, 1047
\bibitem[Etherington(1933)]{Etherington1} Etherington, I. M. H. 1933, Phil. Mag., 15, 761
\bibitem[Etherington(2007)]{Etherington2} Etherington, I. M. H. 2007, Gen. Rel. Grav., 39, 1055
\bibitem[Foreman-Mackey \& Hogg(2013)]{Foreman13}Foreman-Mackey, D., Hogg, D. W., Lang, D., \& Goodman, J. 2013, PASP, 125, 306
\bibitem[Fujimoto et al.(2019)] {Fujimoto19} Fujimoto, S., et al. 2019, ApJ, 887, 107
\bibitem[Holanda et al.(2013)]{Holanda13}Holanda, R. F. L., et al. 2013, JCAP, 1304, 027
\bibitem[Jaeckel \&Ringwald(2010)]{Jaeckel10}Jaeckel, J., \& Ringwald, A. 2010, Ann. Rev. Nucl. Part. Sci. 60, 405
\bibitem[Jesus et al.(2017)]{Jesus17}Jesus, J. F., Holanda, R. F. L., \& Dantas, M. A. 2017, General Relativity and Gravitation, 49, 150
\bibitem[Kawamura et al.(2006)]{Kawamura06}Kawamura, S.,  Nakamura, T.,  Ando., M. et al., 2006, CQG, 23, 125
\bibitem[Kawamura et al.(2019)]{Kawamura19}Kawamura, S.,  Nakamura, T.,  Ando., M. et al., 2019, IJMPD, 28, 1845001
\bibitem[Kawamura et al. (2011)]{Kawamura11}Kawamura, S., et al. 2011, CGQ, 28, 094011
\bibitem[Kessler \& Scolnic(2017)]{Kessler17}Kessler, R., \& Scolnic, D. 2017, ApJ, 836, 56
\bibitem[Kidder et al.(1993)]{Kidder93} Kidder., L. E.,  Will, C. M., \& Wiseman, A. G. 1993, PRD, 47, 3281
\bibitem[Li et al.(2013)]{Li13} Li, Z., et al. 2013, PRD, 87, 103013
\bibitem[Li \& Draine(2001)] {Li01} Li, A. \& Draine, B. T. 2001, ApJ, 554, 778
\bibitem[Liao et al.(2013)]{Liao13} Liao, K., et al. 2013, PLB, 718, 1166
\bibitem[Liao et al.(2015a)]{Liao15} Liao, K., et al. 2015a, PRD, 92, 123539
\bibitem[Liu et al.(2020a)]{Liu20a} Liu, Y. T., et al. 2020a, ApJ, 901, 129
\bibitem[Liu et al.(2020b)]{Liu20b} Liu, T. H., et al. 2020b, ApJ, 899, 71
\bibitem[Liu et al.(2020c)]{Liu20c} Liu, T. H., et al. 2020c, MNRAS, 496, 708
\bibitem[Ma et al.(2017)]{Ma17} Ma, Y.-B., et al. 2017, EPJC, 77, 891
\bibitem[Ma \& Corasaniti(2018)]{Ma18}Ma, C. \& Corasaniti, P.-S. 2018, ApJ, 861, 124
\bibitem[Ma et al.(2017)]{Ma17} Ma, Y.-B., et al. 2017, EPJC, 77, 891
\bibitem[Ma et al.(2019)]{Ma19} Ma, Y.-B., et al. 2019, ApJ, 887, 163
\bibitem[Maggiore(2008)]{Maggiore08} Maggiore, M. Gravitational Waves, 2008, Oxford University Press, New York
\bibitem[Mathis(1990)] {Mathis90} Mathis, J. S. 1990, ARA\&A, 28, 37
\bibitem[Melia(2019)]{Melia19} Melia, F. 2019, MNRAS, 489, 517
\bibitem[Melnick et al.(1987)]{Melnick18} Melnick, J., Moles, M., Terlevich, R., Garcia-Pelayo, J.-M., 1987, MNRAS, 226, 849
\bibitem[Menard et al.(2010a)]{Menard10a} Menard, B., Scranton, R., Fukugita, M., \& Richards, G. 2010a, MNRAS, 405, 1025
\bibitem[Menard et al.(2010b)]{Menard10b} Menard, B., Kilbinger, M., \& Scranton, R. 2010b, MNRAS, 406, 1815
\bibitem[Messenger \& Read(2012)]{Messenger12} Messenger, C. \& Read, J. 2012, PRL, 108, 091101
\bibitem[Messenger \& Takami(2014)]{Messenger14} Messenger, C., Takami, K., Gossan, S., Rezzolla, L., \& Sathyaprakash, B. S. 2014, PRX 4, 041004
\bibitem[More et al.(2009)]{More09} More, S., et al. 2009, ApJ, 696, 1727
\bibitem[Morgan et al.(2014)]{Morgan14} Morgan, A. N., et al. 2014, MNRAS, 440, 1810
\bibitem[Nair et al.(2012)]{Nair12} Nair, R., Jhingan, S., \& Jain, D. 2012, JCAP, 12, 028
\bibitem[Nishizawa et al.(2010)]{Nishizawa10} Nishizawa, A., Taruya, A., \& Kawamura, S. 2010, PRD, 81, 104043
\bibitem[Paris et al.(2017)]{Paris17} Paris, I., et al. 2017, A\&A, 597, 79
\bibitem[Perlmutter et al.(1999)]{Perlmutter99} Perlmutter, S., et al. 1999, ApJ, 517, 565
\bibitem[Pi\'{o}rkowska-Kurpas et al.(2020)] {Ola20} Pi\'{o}rkowska-Kurpas, A., et al. 2020, ApJ, submitted [arXiv:2005.08727]
\bibitem[Planck Collaboration(2018)]{Planck18} Aghanim, N., Akrami, Y., Ashdown, M., et al. [Planck Collaboration] 2018, arXiv:1807.06209
\bibitem[Plionis et al.(2011)]{Plionis2011} Plionis, M., et al. 2011, MNRAS, 416, 2981
\bibitem[Prochaska \& Herbert-Fort(2004)] {Prochaska04} Prochaska, J. X. \& Herbert-Fort, S. 2004, PASP, 116, 622
\bibitem[Qi et al.(2018)]{Qi18} Qi, J.-Z., et al. 2018, RAA, 18, 66
\bibitem[Qi et al.(2019a)]{Qi19a} Qi, J. Z., et al. 2019a, PRD, 100, 023530
\bibitem[Qi et al.(2019b)]{Qi19b} Qi, J. Z., et al. 2019, PDU, 26, 100338
\bibitem[Rao et al.(2006)] {Rao06} Rao, S. M., Turnshek, D. A., \& Nestor, D. B. 2006, ApJ, 636, 610
\bibitem[Ratra \& Peebles(1988)]{Ratra88}  Ratra, B., \& Peebles, P. E. J. 1988, PRD, 37, 3406
\bibitem[Rest \& Scolnic (2014)]{Rest14}Rest, A., Scolnic, D., Foley, R. J., et al. 2014, ApJ, 795, 44
\bibitem[Riess et al.(1998)]{Riess98} Riess, A.~G., et al. 1998, AJ, 116, 1009
\bibitem[Risaliti \& Lusso(2015)]{Risaliti2015}Risaliti, G., \& Lusso, E. 2015, ApJ, 815, 33
\bibitem[Risaliti \& Lusso(2019)]{Risaliti2018}Risaliti, G., \& Lusso, E. 2019, Nature Astronomy, 3, 272
\bibitem[Rosen et al.(2016)]{Rosen2016}Rosen, S. R., et al. 2016, A\&A, 590, 1
\bibitem[Sathyaprakash et al.(2010)]{Sathyaprakash2010} Sathyaprakash, B., et al. 2010, CQG, 27, 215006
\bibitem[Schutz (1986)]{Schutz86}Schutz, B. F. 1986, Nature, 323, 310
\bibitem[Scolnic et al.(2014)]{Scolnic14}Scolnic, D., Rest, A., Riess, A., et al. 2014, ApJ, 795, 45
\bibitem[Scolnic et al.(2018)]{Scolnic18} Scolnic, D., et al. 2018, ApJ, 859, 101
\bibitem[Schneider et al.(2001)]{Schneider01}Schneider, R., et al. 2001, MNRAS, 324, 797
\bibitem[Seto et al.(2001)]{Seto01} Seto, N., Kawamura, S., \& Nakamura, T. 2001, PRL, 87, 221103
\bibitem[Shen et al.(2011)]{Shen11} Shen, Y. et al. 2011, ApJS, 194, 45
\bibitem[Siegel et al.(2005)]{Siegel2005} Siegel, E. R., et al. 2005, MNRAS, 356, 1117
\bibitem[Terlevich \& Melnick (1981)]{Terlevich81}Terlevich, R. \& Melnick, J., 1981, MNRAS, 195, 839
\bibitem[Terlevich et al.(2015)]{Terlevich2015} Terlevich, R., et al. 2015, MNRAS, 451, 3001
\bibitem[Tolman(1930)]{Tolman30} Tolman, R. C. 1930, Proc. Natl. Acad. Sci., 16, 511
\bibitem[Vavry\v{c}uk(2019)]{Vavrycuk19} Vavry\v{c}uk, V. 2019, MNRAS, 489, L63
\bibitem[Vavry\v{c}uk \& Kroupa(2020)]{Vavrycuk20} Vavry\v{c}uk, V, \& Kroupa, P. 2020, MNRAS, 497, 378
\bibitem[Wang et al.(2017)]{Wang17} Wang, G.-J., Wei, J.-J., Li, Z.-X., Xia, J.-Q., \& Zhu, Z.-H. 2017, ApJ, 847, 45
\bibitem[Watson et al.(2015)] {Watson15} Watson, D., et al. 2015, Nature, 519, 327
\bibitem[Wei et al.(2016)]{JunJie2016} Wei, J.-J., Wu, X.-F., \& Melia, F. 2016, MNRAS, 463, 1144
\bibitem[Wei(2019)]{Wei19} Wei, J.-J. 2019, ApJ, 876, 1
\bibitem[Wu et al.(2020)]{Wu20} Wu, Y., et al. 2020, 888, 113
\bibitem[Xie et al.(2015)]{Xie15} Xie, X., Shen, S., Shao, Z., \& Yin, J. 2015, ApJ, 802, L16
\bibitem[Xu et al.(2018)]{Xu18} Xu, T., et al. 2018, JCAP, 06, 042
\bibitem[Yagi \& Seto (2011)]{Yagi11}Yagi, K. \& Seto, N. 2011, PRD, 83, 044011
\bibitem[Zhao et al.(2011)] {Zhao11} Zhao, W., Van Den Broeck, C., Baskaran, D., \& Li, T. 2011, PRD, 83, 023005
\bibitem[Zhang et al.(2020)]{Zhang20} Zhang, S., et al. 2020, IJMPD, in pess [arXiv:2009.04204]
\bibitem[Zheng et al.(2020)]{Zheng20} Zheng, X., et al. 2020, ApJ, 892, 103












\end{thebibliography}
\end{document}